# On the power-law *q*-distribution function based on the probabilistically independent postulate in nonextensive statistics

Du Jiulin[*]

*Department of Physics, Shool of Science, Tianjin Universitu, Tianjin 300072, China*

**Abstract** We deal with the power-law *q*-distribution functions, so-called *q*-exponentials in nonextensive statistics. The system considered is a many-body Hamiltonian system with arbitrary interacting potentials. We find that the usual form of power-law *q*-distribution function employed in nonextensive statistics for the system may be not to stand by the probabilistically independent postulate and it only represents a dynamical isothermal situation. The probabilistically independent postulate is validated by the nonextensivity (or pseudoadditivity) in nonextensive statistics and thus the usual forms of power-law *q*-distribution function have to be modified. In this letter, we present a new power-law *q*-distribution based on the probabilistically independent postulate, which can represent nonequilibrium stationary state of the system away from equilibrium and with the nonextensive energy.

---

[*] Email address: jldu@tju.edu.cn

Nonextensive statistics is being recognized as one reasonable generalization to Boltzmann-Gibbs statistics. The probabilistic independence is one of the basic postulates in nonextensive statistics. And based on this postulate, nonextensive statistics has been investigated and developed, and it has been made very wide applications in many interesting scientific fields for complex systems since the *q*-entropy published by Tsallis in 1988 (see Ref.[1-3] and http://tsallis.cat.cbpf.br/biblio.htm). Generally speaking, it has been known that nonextensive statistics is founded on the basis of Tsallis *q*-entropy and the probabilistically independent postulate. By the maximum *q*-entropy principle with some of the given conditions, one obtained the power-law *q*-distribution, the so-called *q*-exponentials in nonextensive statistics. Based on the probabilistically independent postulate, one could find the nonextensivity (also often called the pseudoadditivity or nonadditivity) both of the *q*-entropy and the energy [4]. However, when nonextensive statistics is studied and it is made the applications in many-body Hamiltonian systems, one has actually discarded the nonextensivity (or the pseudoadditivity) of the energy so far. And the currently employed form of the power-law *q*-distribution function in nonextensive statistics may be also not to stand by the probabilistically independent postulate, neglecting the nonextensivity of the energy, and it seems to only stand for an isothermal distribution in many situations. On the other hand, from the second law of thermodynamics, e.g. $dU = TdS$ (if the volume is fixed), we can find that it is hard to image that the entropy is nonextensive but the energy is extensive. To solve the inconsistency problem, in this letter, we try to suggest a new power-law *q*-distribution function for nonextensive statistics that the *q*-exponential distribution of a Hamiltonian system is written as factorization of the *q*-exponentials with composite energies.

The *q*-entropy is proposed by Tsallis as a generalization of Boltzmann-Gibbs statistics with the mathematical expression [4],

$$S_q = -k \sum_i p_i^q \ln_q p_i \qquad (1)$$

where *k* is Boltzmann constant, the set {$p_i$} are the probabilities of the microscopic configurations {*i*} of the system under consideration, and the index *q* is the nonextensive parameter whose difference from unity is thought of measuring the degree of the nonextensivity for the system. The function, the $q$-logarithm is defined as

$$\ln_q x \equiv \frac{x^{1-q}-1}{1-q}, \quad (x>0; \ln_{q=1} = \ln x). \tag{2}$$

Its inverse function, the $q$-exponential, is

$$\exp_q x \equiv [1+(1-q)x]^{\frac{1}{1-q}}, \quad (\exp_{q=1} x = \exp x), \tag{3}$$

if $1+(1-q)x >0$ and by $\exp_q x = 0$ otherwise. Thus, the probability of a system at the value $x$ reads $p \sim \exp_q x$, being a power-law $q$-distribution function. By extremizing the $q$-entropy Eq.(1) with the given constraint conditions [4, 5], i.e. the energy and $\sum_i p_i = 1$, one finds the general form of the power-law $q$-distribution in nonextensiv statistics,

$$p_i \sim [1-(1-q)\beta H^{(i)}]^{1/1-q} = \exp_q(-\beta H^{(i)}). \tag{4}$$

where $\beta = 1/kT$ is the Lagrange parameter and $H^{(i)}$ is the energy at $i$th configuration. Obviously, when we take $q=1$, all of the formulae in the scheme return to be those in Boltzmann-Gibbs statistics.

The so-called probabilistically independent postulate is that, if the probability of a system $A$ at its $i$th configuration with the value $x_i$ is $p_i^A \sim \exp_q x_i$ and a system $B$ at its $j$th configuration with the value $x_j$ is $p_j^B \sim \exp_q x_j$, and they are probabilistically independent, then the joint probability of the composite system $A \oplus B$ at its configuration $(i, j)$ with the value $(x_i+x_j)_q$ is

$$p_{ij}^{A \oplus B}(x_i, x_j) \sim \exp_q(x_i + x_j)_q$$

and

$$p_{ij}^{A \oplus B} = p_i^A p_j^B \sim (\exp_q x_i)(\exp_q x_j). \tag{5}$$

More generally, the postulate can be written as the joint probability to be

$$p(\{x_i\}) = \prod_i p_i \sim \prod_i \exp_q x_i, \tag{6}$$

where the quantity $x_i$ is nonextensive (or pseudoadditive) so as to validate Eq.(5) and Eq.(6).

With this basic postulate, for a composite system with $N$ subsystems, their entropy and energy being $S_q^{(i)}$ and $H^{(i)}$, $i=1, 2, \ldots N$, respectively, we can find the nonextensivity (or pseudoadditivity) both of the $q$-entropy and the energy, expressed by

$$\ln[1+(1-q)S_q/k] = \sum_{i=1}^{N} \ln[1+(1-q)S_q^{(i)}/k], \tag{7}$$

and

$$\ln[1+(1-q)\beta H_q] = \sum_{i=1}^{N} \ln[1+(1-q)\beta H^{(i)}]. \tag{8}$$

Further they can be written as

$$1+(1-q)S_q/k = \prod_{i=1}^{N}[1+(1-q)S_q^{(i)}/k], \tag{9}$$

and

$$1+(1-q)\beta H_q = \prod_{i=1}^{N}[1+(1-q)\beta H^{(i)}], \tag{10}$$

where $S_q$ and $H_q$ are the total $q$-entropy and the total $q$-energy of the composite system. Eq.(9) and Eq.(10) are the two basic relations to coexist in nonextensive statistics. If $N=2$, they become the following relations familiar to us, viz.

$$S_q^{A \oplus B} = S_q^{(A)} + S_q^{(B)} + (1-q)k^{-1}S_q^{(A)}S_q^{(B)}, \tag{11}$$

and

$$H_q^{A \oplus B} = H^{(A)} + H^{(B)} + (1-q)\beta H^{(A)}H^{(B)}, \tag{12}$$

The probabilistic independence is as a basic postulate for nonextensive statistics, which is known to hold not only for independent noninteracting systems but also for the interacting systems. On the other hand, if the postulate of probabilistic independence was incorrect, many in nonextensive statistics including Eq.(11) and Eq.(12) should need to be reconsidered.

We consider a many-body Hamiltonian system with $N$ identical particles of mass $m$ and arbitrary interacting potentials. The Hamiltonian of the system reads

$$H(\mathbf{r},\mathbf{p}) = \sum_{i=1}^{N} \frac{\mathbf{p}_i^2}{2m} + V(\mathbf{r}_1,\ldots,\mathbf{r}_N), \tag{13}$$

where $(\mathbf{r}, \mathbf{p})=(\mathbf{r}_1, \ldots \mathbf{r}_N; \mathbf{p}_1, \ldots \mathbf{p}_N)$, $V(\mathbf{r}_1, \ldots \mathbf{r}_N)$ is the total interacting potential composed of $N(N-1)/2$ pair interactions of particles in the system, written as

$$V(\mathbf{r}_1, \ldots \mathbf{r}_N) = \sum_{i<j}^{N} u(|\mathbf{r}_i - \mathbf{r}_j|). \tag{14}$$

If one wants to apply the power-law $q$-distribution Eq.(4) to the above system, usually in many situations, e.g. in [6-9], the $q$-distribution function is written directly by inserting the total Hamiltonian of the system Eq.(13) in Eq.(4),

$$\begin{aligned} p_q &\sim [1-(1-q)\beta H(\mathbf{r},\mathbf{p})]^{1/1-q} \\ &= \{1-(1-q)\beta [\sum_{i=1}^{N} \frac{\mathbf{p}_i^2}{2m} + V(\mathbf{r}_1,\ldots,\mathbf{r}_N)]\}^{1/1-q} \end{aligned} \tag{15}$$

In the present point of view, this kind of $q$-distributions employed in nonextensive statistics is not to stand by the probabilistically independent postulate, and it neglects the nonextensivity of the energy in Eq.(10). In particular, from the dynamical analyses of stationary state solutions of a Fokker-Planck equation, it is known that the $q$-distribution with the shape such as Eq.(15) only represents a dynamical isothermal situation for the Hamiltonian system with an arbitrary potential [10]. Namely, for the $q$-distribution Eq.(15), there is always the equation,

$$\frac{d\beta}{dr} = 0, \tag{16}$$

which only stands for an isothermal distribution describing the thermal equilibrium state. When one employed the $q$-distribution Eq.(15) as the power-law distribution to study a Hamiltonian system, one might not be conscious that the distribution function was actually an isothermal one and it contravened the probabilistically independent postulate and thus discarded the nonextensivity of energy. The example was reported recently in the $N$-body simulation for a self-gravitating system with the result that the Tsallis distribution is inconsistency generally with the dark matter halos except the isothermal parts for the polytropic index $n \to \infty$ [11]. However, theoretically, the Tsallis distribution function usual employed in the self-gravitating collisionless system was found to be only an isothermal distribution for any $q \neq 1$ [12].

In order to solve the above problems, we suggest the following scheme. In fact, one can check that the $q$-exponential of a sum cannot be applied *mechanically*, as in

Eq.(15), to the above many-body Hamiltonian system, namely,

$$\exp_q \sum_i x_i \neq \left[1+(1-q)\sum_i x_i\right]^{1/1-q}, \quad (17)$$

but it can be written as the product of the $q$-exponentials based on the probabilistically independent postulate Eq.(6),

$$\exp_q(\sum_i x_i)_q = \prod_i \exp_q x_i = \prod_i [1+(1-q)x_i]^{1/1-q} \quad (18)$$

where the quantity $x_i$ ($i=1,2,…$) is now nonextensive (or pseudoadditive) so as to validate Eq.(18), saying they have a pseudoadditive relation similar to Eq.(10). In terms of Eq.(18), the power-law $q$-distribution of the many-body Hamiltonian system is the factorization of the probabilities, given by

$$p_q^{(N)} \sim \prod_{i=1}^N \left[1-(1-q)\beta\frac{\mathbf{p}_i^2}{2m}\right]^{1/1-q} \times [1-(1-q)\beta V(\mathbf{r}_1,...\mathbf{r}_N)]^{1/1-q}$$

$$= \prod_{i=1}^N \left[1-(1-q)\beta\frac{\mathbf{p}_i^2}{2m}\right]^{1/1-q} \times \prod_{i<j}^N [1-(1-q)\beta u(|\mathbf{r}_i-\mathbf{r}_j|)]^{1/1-q}. \quad (19)$$

For a one-body Hamiltonian system, $N=1$, the power-law $q$-distribution (e.g. one dimension system under spherical symmetry) is

$$p_q^{(N=1)} \sim [1-(1-q)\beta p^2/2m]^{1/1-q} \times [1-(1-q)\beta V(r)]^{1/1-q}. \quad (20)$$

For the above $q$-distributions, the $q$-entropy and the energy of the many-body Hamiltonian system are both nonextensive (or pseudoadditive). We can educe the pseudoadditive relation between the total energy and its independent parts by using Eq.(10). For the $q$-distribution (20), for example, we deduce the total energy of the system,

$$H_q^{(N=1)} = \frac{p^2}{2m} + V(r) + (1-q)\beta V(r)\frac{p^2}{2m}. \quad (21)$$

For the $q$-distribution (19), we have

$$1+(1-q)\beta H_q^{(N)} = \prod_{i=1}^N [1+(1-q)\beta \mathbf{p}_i^2/2m] \times \prod_{i<j}^N [1+(1-q)\beta u(|\mathbf{r}_i-\mathbf{r}_j|)]. \quad (22)$$

Mathematically, this kind of pseudoadditivity for energy validates the postulate Eq.(18). Accordingly, Eq.(19) and Eq.(20) can be written simply as,

$$p_q^{(N)} \sim \left[1 - (1-q)\beta H_q^{(N)}\right]^{1/1-q}, \tag{23}$$

$$p_q^{(N=1)} \sim \left[1 - (1-q)\beta H_q^{(N=1)}\right]^{1/1-q}. \tag{24}$$

Thus, the old $q$-distribution Eq.(15) should be replaced by the new $q$-distribution Eq.(23).

The dynamics of the new power-law $q$-distribution Eq.(19) and the physical meaning of $q$ can be studied from the stationary state solutions of Fokker-Planck equation [10,13]. By a scheme of analyses, it is shown that the new $q$-distribution can be as a statistical description for the nonequilibrium dynamics of the Hamiltonian system with an arbitrary potential $V(r)$ [14]. Namely, for the $q$-distribution Eq. (20) or Eq.(24) we can get the following relation,

$$\frac{d\beta}{dr} = (1-q)\beta^2 m \frac{dV}{dr}. \tag{25}$$

Or it is written by

$$1 - q = -\frac{k}{m}\frac{dT}{dr} \bigg/ \frac{dV}{dr}, \tag{26}$$

which shows that Eq.(20) is the nonequilibrium stationary distribution satisfying the relation (25) if $q$ is different from unity. The nonextensive parameter $q$ is equal to unity if and only if $d\beta/dr = 0$ (i.e. $dT/dr = 0$). The corresponding Langevin equation for the dynamics shows that this new power-law $q$-distribution is therefore a potential statistical distribution for dark matters. The reader might also be interested in some recent applications of nonextensive statistical mechanics, where the physical meanings of the nonextensive parameter $q \neq 1$ is introduced to astrophysical systems and plasmas [12,15-18]. Works related to factorizations of the $q$-distribution were also reported recently, e.g. in [12,14, 19].

In conclusive remarks, if the probabilistically independent postulate is correct one in nonextensive statistics, the power-law $q$-distribution for a many-body Hamiltonian system has to be modified, or we may adopt the form such as Eq.(23), in which the $q$-entropy and the energy of the system are both pseudo-additive or nonextensive. This kind of pseudoadditivity or nonextensivity for the energy validates the postulate. In reverse, if the probabilistically independent postulate was incorrect one in nonextensive statistics, then many in nonextensive statistics and its applications including the nonextensivity (or pseudoadditivity ) of *q*-entropy should be reconsidered.

**Acknowledgement**

This work was supported by the National Natural Science Foundation of China under Grant No.10675088. I thank S. Abe and C. Beck for useful discussions.